\documentclass[aps,twocolumn,preprintnumbers,superscriptaddress,nofootinbib]{revtex4}
\usepackage{amsmath} \usepackage{graphicx} \usepackage{amsfonts}
\usepackage{array} \usepackage{amsthm} \usepackage{bm}
\usepackage{palatino} \usepackage{mathpazo} 

\usepackage[breaklinks]{hyperref}
\usepackage{color}

\usepackage{epstopdf}

\newcommand{\nn}{\nonumber}
\newcommand{\be}{\begin{equation}}
\newcommand{\ee}{\end{equation}}
\newcommand{\ba}{\begin{eqnarray}}
\newcommand{\ea}{\end{eqnarray}}
\newcommand{\bal}{\begin{align}}
\newcommand{\eal}{\end{align}}

\newcommand{\dd}{{\rm d}}

\newcommand{\al}{\alpha}

\newcommand{\bt}{\beta}

\newcommand{\ga}{\gamma}

\newcommand{\si}{\sigma}
\newcommand{\ta}{\theta}

\newcommand{\vp}{\varphi}

\newcommand{\bw}{\begin{widetext}}
\newcommand{\ew}{\end{widetext}}

\begin{document}

\title{Strong Gravitational Lensing by a Charged Kiselev Black Hole}

\author{Mustapha Azreg-A\"{\i}nou}\email{azreg@baskent.edu.tr}
\affiliation{Engineering Faculty, Ba\c{s}kent University, Ba\u{g}l\i ca Campus, Ankara, Turkey}

\author{Sebastian Bahamonde}
\email{sebastian.beltran.14@ucl.ac.uk}
\affiliation{Department of Mathematics, University College London,
    Gower Street, London, WC1E 6BT, UK}

\author{Mubasher Jamil}
\email{mjamil@sns.nust.edu.pk}
\affiliation{Department of Mathematics, School of Natural
Sciences (SNS), National University of Sciences and Technology
(NUST), H-12, Islamabad, Pakistan}

\begin{abstract}
We study the gravitational lensing scenario
where the lens is a spherically symmetric charged black hole (BH) surrounded by
quintessence matter. The null geodesic equations in the curved
background of the black hole are derived. The resulting
trajectory equation is solved analytically via perturbation and series methods
for special choice of parameters and the distance of the closest
approach to black hole is calculated. We also derive the lens
equation giving the bending angle of light in the curved background.
In the strong field approximation, the solution of the lens equation
is also obtained for all values of the quintessence parameter $w_q$. For all
$w_q$, we show that there are no stable closed null orbits and that corrections to the deflection angle for the Reissner-Nordstr\"om black hole when the observer and the source are at large, but finite, distances from the lens do not depend on the charge up to the inverse of the distances squared. A part of the present work,
analyzed however with a different approach, is the extension
of {\it Phys. Rev. D \textbf{92}, 084042 (2015)} where the uncharged case has been treated.\\
\textbf{Keywords}: Gravitational lensing; charged black hole; quintessence; null-geodesics
\end{abstract}

\pacs{}

\maketitle

\section{Introduction}
It is predicted by General Relativity (GR) that in  the presence of
a mass distribution, the light is deflected. However, it was not
entirely a new prediction by Einstein, in fact, Newton had obtained
a similar result by a different set of assumptions. In 1936,
Einstein \cite{Ein} noted that if a star (lens), the background star
(source) and the observer are highly aligned then the image obtained
by the deflection of light of a background star due to another star
can be highly magnified. He also noted that optical telescopes at
that time were not enough capable to resolve the angular separation
between images.

In 1963, the discovery of quasars at high redshift gave the actual
observation to the gravitational lensing effects. Quasars are
central compact light emitting regions which are extremely luminous. When a galaxy appears between the quasar and the observer,
the resulting magnification of images would be large and hence well
separated images are obtained. This effect was named macro-lensing. The first
example of gravitational lensing was discovered (the quasar QSO
0957+561) in 1979 \cite{Walsh}.

The weak field theory of gravitational lensing is based on the first
order expansion of the smallest deflection angle. It has been
developed by several authors such as Klimov \cite{Kli}, Liebes
\cite{Lie}, Refsdal \cite{Ref}, Bourassa \cite{Bou,Bou1,Bou2}, and
Kantowski \cite{Kan}. They were succeeded in explaining astronomical
observations up to now (for more detailed see \cite{Schne}).

Due to a highly curved spacetime by a black hole (BH), the weak
field approximation is no longer valid. Ellis and Virbhadra obtained
the lens equation by studying the strong gravitational fields
\cite{Virbh}. They analyzed the lensing of the Schwarzchild BH with an
asymptotically flat metric. They found two infinite sets of faint
relativistic images with the primary and secondary images. Fritelli
{\it et al.} \cite{Frit} obtained exact lens equation and they
compared them with the results of Virbhadra and Ellis. By using the
strong field approximation, Bozza {\it et al.} \cite{Bozza} gave an
analytical expressions for the magnification and positions of the
relativistic images.

From recent observational measurements, we can see that our Universe
is dominated by a mysterious form of energy called ``Dark Energy".
This kind of energy is responsible for the accelerated expansion of
our Universe \cite{Pee, Wan}. Dark energy acts as a repulsive
gravitational force so that usually it is modeled as an exotic
fluid. One can consider a fluid with an equation of state in which
the state parameter $w(t)$ depends on the ratio of the pressure
$p(t)$ and its energy density $\rho(t)$ such as: $w(t) =
\frac{p(t)}{\rho(t)}$. So far, wide variety of dark energy models
with dynamical scalar fields have been proposed as alternative
models to the cosmological constant. Such scalar field models
include: quintessence \cite{Kis,Kis1,Kis2,Shau}, k-essence \cite{Yand},
quintom \cite{Guo, Xia}, phantom dark energy \cite{Mart} and
others.

Quintessence is a candidate of dark energy which is represented by
an exotic kind of scalar field that is varying with respect to the
cosmic time. The solution for a spherically symmetric spacetime
geometry surrounded by a quintessence matter was derived by Kiselev
\cite{Kis}. There are few works focused on studying the Kiselev
black hole (KBH).
Thermodynamics and phase transition of the Reissner-Nordstr\"{o}m BH
surrounded by quintessence are given in \cite{Kis1,Kis2,Tho}. The thermodynamics of the Reissner–Nordström–de Sitter black hole surrounded by quintessence has been investigated by one of us~\cite{Kis2} and has led to the notion of two thermodynamic volumes.
The properties
of charged BH surrounded by the quintessence were studied in
\cite{Fer, Kis2}. New solutions that generalize the Nariai horizon to
asymptotically de Sitter-like solutions surrounded by quintessence have been determined in~\cite{Kis2}. The detailed study of the photon trajectories around the
charged BH surrounded by the quintessence is given in \cite{FMR}.
Recently, A. Younas and collaborators worked on the strong
gravitational lensing by Schwarzschild-like BH surrounded by
quintessence \cite{Skv}.

In the following paper, we will extend that
work by adding a charge $Q$ (charged KBH). We will consider the
lensing phenomenon only for the case of non-degenerate horizons. By
computing the null geodesics, we examine the behavior of light
around a charged KBH. We analyze the circular orbits (photon region)
for photons. Furthermore, we observe how both the quintessence and
the charge parameters affects the light trajectories of massless
particles (photons), when they are strongly deflected due to the charged KBH. We will not restrict the investigation to the analytically trackable cases $w_{q}=-1/3$ and $w_{q}=-2/3$, as some works did~\cite{Fer,FMR,SB}; rather, we will consider the full range of the quintessence parameter $w_q$ and we will rely partly on the work done by one of us~\cite{Kis2}.

The paper is structured as follows: In Sec.~II, we study the charged
KBH geometry and we derive the basic equations for null geodesics. Additionally, in that section we write down the basic equations for null geodesics in
charged Kiselev spacetime along with the effective potential and the
horizons. In Sec.~III, the analytical solution of the trajectory
equation is obtained via perturbation technique. Sec.~IV is devoted
to the study of the lens equation to derive the bending angle. The
strong field approximation of the lens equation is discussed as well.
Finally, we provide a conclusion in Sec.~VI. Throughout this paper, we adopt the
natural system of units where $c = G = 1$ and the metric convention
$(+,-,-,-)$.

\section{Basic equations for null geodesics in charged Kiselev spacetime}
The geometry of a charged KBH surrounded by quintessence  is given by \cite{Kis}
\begin{equation}\label{metric}
ds^{2}=f(r)dt^{2}-\frac{1}{f(r)}dr^{2}-r^{2}d\theta^{2}-r^{2}\sin^{2}\theta d\phi^{2},
\end{equation}
where
\begin{equation}\label{2.1}
f(r)=1-\frac{2M}{r}-\frac{\sigma}{r^{3w_{q}+1}}+\frac{Q^{2}}{r^{2}}.
\end{equation}
Here, $M$ is the mass of the BH, $w_{q}$ is the quintessence state parameter (having range between $-1\leq w_{q} <-1/3$), $\sigma$ is a positive normalization factor and $Q$ is the charge of the BH. The equation of state for the quintessence matter with isotropic negative pressure $p_{q}$ is linear of the form
\begin{equation}
p_{q}=w_{q}\rho_{q}<0,
\end{equation}
where $\rho_{q}$ is the energy density given by (taking $G=\hbar=1$)
\begin{equation}
\rho_{q}=-\frac{3w_{q}\sigma}{8\pi r^{3(1+w_{q})}}>0.
\end{equation}
For a detailed metric derivation and a discussion of its properties, we refer the reader to the original paper by Kiselev~\cite{Kis}. For a further discussion see~\cite{Kis1,Kis2}. Note that not all the values of $w_q$ are manageable to find analytically solutions to the trajectory equation (See Sec. II). However, the cosmological constant case, corresponding to $w_q=-1$, and the case $w_q=-2/3$ are relatively simple.

\subsection{Horizons in charged Kiselev black hole}
In order to study the trajectories of photons near the space-time (\ref{metric}), one has to understand where the horizons are located for the charged KBH. In order to find the horizons, we require $f(r)=0$, which depends on four parameters $M,Q^2,w_{q}$ and $\si$. It has become custom to fix $M,Q^2$ and $w_{q}$~\cite{Kis1,Kis2} and investigate the properties of these BHs upon constraining the values of $\si$ in terms of $M,Q^2$ and $w_{q}$. We proceed the same way in this work.
\begin{figure*}
\centering
\includegraphics[width=0.47\textwidth]{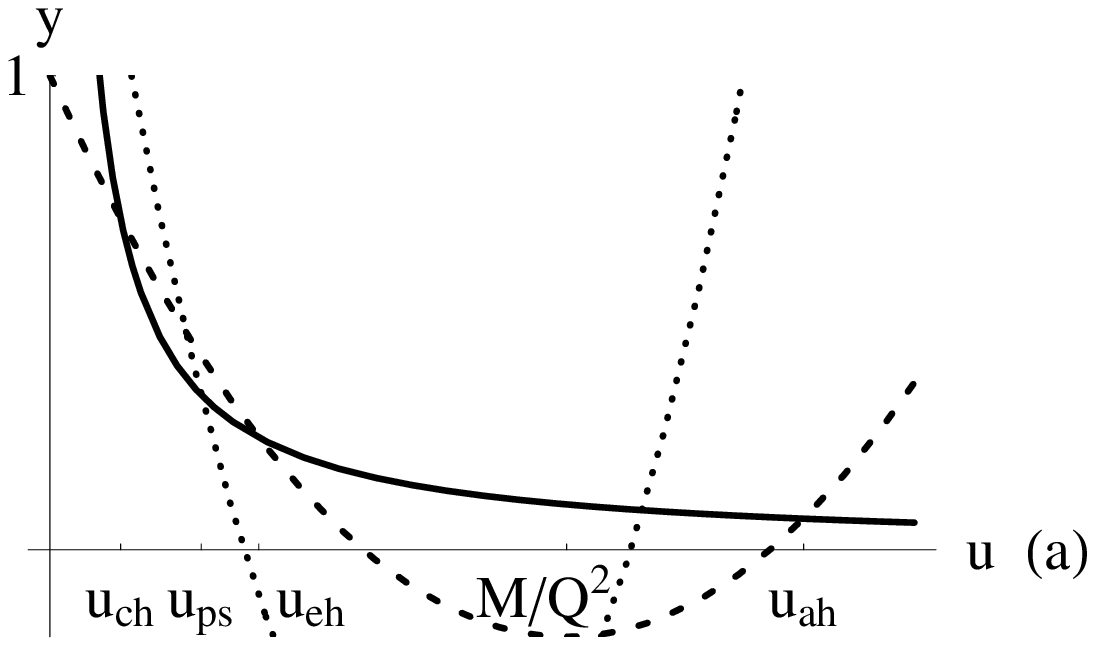}  \includegraphics[width=0.47\textwidth]{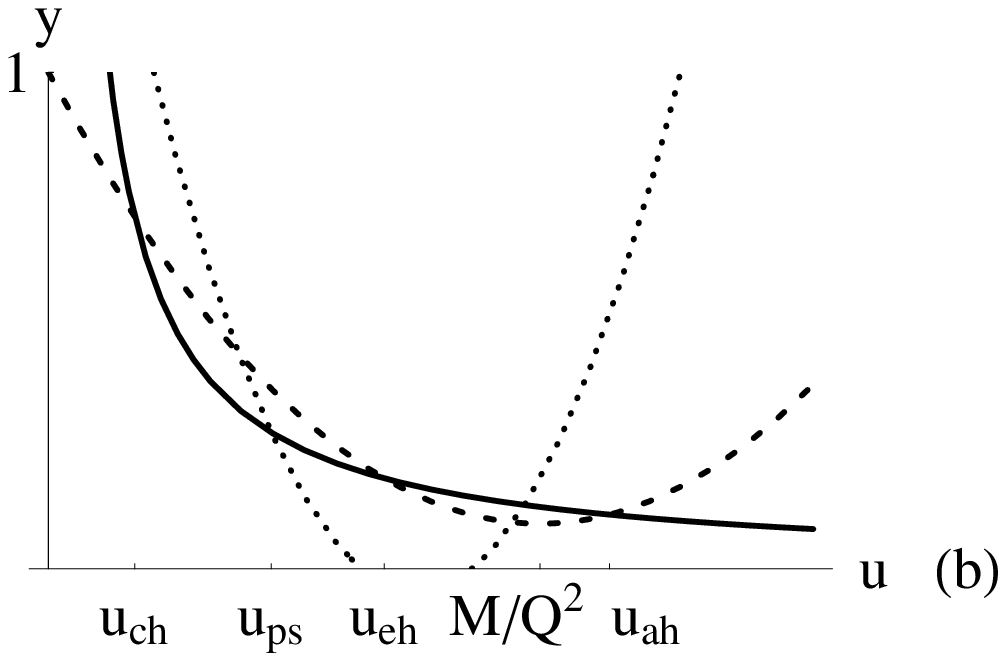} \\
\caption{\footnotesize{Plots of $y=1-2Mu+Q^2 u^2$ (dashed line), $y=\si u^{3w_q+1}$ (continuous line), and $y=(2-6Mu+4Q^2 u^2)/[3(w_q+1)]$ (dotted line) versus $u\equiv 1/r$ for (a) $Q^2/M^2\leq 1$, $-1\leq w_q<-1/3$, and $\si<\si_1$~\eqref{2.3a}; (b) $Q^2/M^2>1$, $-1\leq w_q<-1/3$, and $\si_2<\si<\si_1$~\eqref{2.3b}. Here $u_{ch}=1/r_{ch}$, $u_{eh}=1/r_{eh}$, $u_{ah}=1/r_{ah}$. The points of intersection of the dashed parabola and the continuous line provide the locations of the three horizons~\eqref{2.5}: ($u_{ch},u_{eh},u_{ah}$). The point of intersection of the dotted parabola and the continuous line provides the only local maximum value~\eqref{e1b} of the potential $V_\text{eff}$~\eqref{e1} for $u_{ch}<u<u_{eh}$, which is the location of the photon sphere: $u_{ps}$.}}\label{Fig1}
\end{figure*}

We are only interested in the case where the BH has three distinct horizons: a cosmological horizon $r_{ch}$, an event horizon $r_{eh}$, and an inner horizon $r_{ah}$ with $r_{ch}>r_{eh}>r_{ah}$ and
\begin{align}
&f<0\quad\text{ for }\quad 0<r<r_{ah},\nn\\
\label{2.2}&f>0\quad\text{ for }\quad r_{eh}<r<r_{ch},\\
&f<0\quad\text{ for }\quad r>r_{ch}.\nn
\end{align}
The photon paths are all confined in the region
\begin{equation}
r_{eh}\leq r\leq  r_{ch},
\end{equation}
where $f\geq 0$.

For $M,Q^2$ and $w_{q}$ fixed, the constraints for having three positive distinct roots of $f(r)=0$ depend on the ratio $Q^2/M^2$. There will be three distinct horizons if~\cite{Kis2}
\begin{align}
\label{2.3a}&\frac{Q^2}{M^2}\leq 1\quad\text{ and }\quad \si<\si_1\equiv \frac{2(Q^2u_1-M)}{(3w_q+1)u_1^{3w_q}},\\
&\text{where}\nn\\
&u_1=-\frac{\sqrt{9w_q^2M^2+(1-9w_q^2)Q^2}+3w_q M}{(1-3w_q)Q^2},\nn
\end{align}
or if~\cite{Kis2}
\begin{align}
\label{2.3b}&\frac{Q^2}{M^2}>1\quad\text{ and }\quad \si_2\equiv \frac{2(Q^2u_2-M)}{(3w_q+1)u_2^{3w_q}}<\si<\si_1,\\
&\text{where}\nn\\
&u_2=\frac{\sqrt{9w_q^2M^2+(1-9w_q^2)Q^2}-3w_q M}{(1-3w_q)Q^2}.\nn
\end{align}
In Ref.~\cite{Kis2} it was shown that under the above constraints~\eqref{2.3a} and~\eqref{2.3b} we have
\begin{align}
\label{2.4a}&u_2>u_1>0,\quad \si_1>\si_2>0,\\
\label{2.4a}&r_{ch}>r_{eh}>\frac{Q^2}{M}>r_{ah}>0.
\end{align}

Introducing the variable $u=1/r$, the horizon equation becomes $f(r)=f(u)=0$, yielding the values of the three horizons. This equation takes the following form ($-2\leq 3w_q+1<0$)
\begin{equation}\label{2.5}
1-2Mu+Q^2 u^2 = \si u^{3w_q+1}.
\end{equation}
Figure~\ref{Fig1}, which is a plot of the parabola $y=1-2Mu+Q^2 u^2$ and the curve $y=\si u^{3w_q+1}$, shows the existence of three distinct horizons for $Q^2/M^2\leq 1$ and $Q^2/M^2>1$. In the remaining part of this work we assume that the constraints~\eqref{2.3a} and~\eqref{2.3b} are satisfied.

\subsection{Equations of motion for a photon }

In the presence of a spherically symmetric gravitational field, we can confine the photon orbits in the equatorial plane by taking $\theta=\pi/2$. Therefore, the Lagrangian for a photon travelling in a charged KBH spacetime will be given by
\begin{equation}
\mathcal{L}=f(r)\dot{t}^{2}-\frac{1}{f(r)}\dot{r}^{2}-r^{2}\dot{\phi}^{2},
\end{equation}
where the dot represents the derivative with respect to the affine parameter $\lambda$ for null geodesics. The Euler-Lagrange equations for null geodesics yield
\begin{eqnarray}
\label{tp1}\dot{t}\equiv\frac{dt}{d\lambda}&=&\frac{E}{f(r)},\\
\label{tp2}\dot{\phi}\equiv\frac{d\phi}{d\lambda}&=&\frac{L}{r^{2}}.
\end{eqnarray}
In the above equations, $E$ and $L$ are constants known as the energy and angular momentum per unit mass. Using the condition for null geodesics $g_{\mu \nu}u^{\mu}u^{\nu}=0$, we obtain the equation of motion for photons
\begin{eqnarray}\label{rdot}
\dot{r}&=&L\sqrt{\displaystyle\frac{1}{b^{2}}-\frac{f(r)}{r^{2}}}, \ \ \ \text{where} \ \ \ b\equiv\Big|\frac{L}{E}\Big|.
\end{eqnarray}
Here, $b$ is the impact parameter which is a perpendicular line to the ray of light converging at the observer from the center of the charged KBH. Further, photons experience a gravitational force in the presence of the gravitational field. This force can be expressed via the effective energy potential which is given by ($\dot{r}^2+V_\text{eff}=E^2$)
\begin{equation}\label{e1}
V_\text{eff}=\frac{L^2}{r^{2}}~f(r)=\frac{L^2}{r^{2}}\Big(1-\frac{2M}{r}-\frac{\sigma}{r^{3w_{q}+1}}+\frac{Q^{2}}{r^{2}}\Big).
\end{equation}
In the left hand side of the above equation, the first term
corresponds to a centrifugal potential, the second term represents
the relativistic correction, the third term is due to the presence
of the quintessence field while the fourth term appears due to the
presence of electric charge. The terms appearing with positive
(negative) signs correspond to repulsive (attractive) force fields.

In terms of $u$, $V_\text{eff}$ reads
\begin{equation}\label{e1a}
V_\text{eff}=L^2(u^2-2Mu^3+Q^2u^4-\sigma u^{3w_q+3}).
\end{equation}
Since $f(u_{eh})=f(u_{ch})=0$ and, by~\eqref{2.2}, $f>0$ for $u_{ch}<u<u_{eh}$ ($u_{ch}=1/r_{ch}$, $u_{eh}=1/r_{eh}$), using~\eqref{e1} we see that $V_\text{eff}(u_{eh})=V_\text{eff}(u_{ch})=0$ and $V_\text{eff}>0$ for $u_{ch}<u<u_{eh}$ too. In the non-extremal case, in which we are interested, this implies that the potential $V_\text{eff}$ may have only an odd number of extreme values between the two horizons $u_{ch}$ and $u_{eh}$; that is, $n+1$ local maxima and $n$ local minima with $n\in \mathbb{N}$. These extreme values are determined by the constraint $dV_\text{eff}/du=0$ which reads ($-2\leq 3w_q+1<0$)
\begin{equation}\label{e1b}
\frac{2}{3(w_q+1)}-\frac{6Mu}{3(w_q+1)}+\frac{4Q^2u^2}{3(w_q+1)}=\si u^{3w_q+1}.
\end{equation}
In absolute value, the slope of the parabola on the lhs of~\eqref{e1b} is larger than that of the parabola on the lhs of~\eqref{2.5} [recall $-1\leq w_q<-1/3$]. Thus, referring to Fig.~\ref{Fig1}, the parabola on the lhs of~\eqref{e1b} intersects the curve $y=\si u^{3w_q+1}$ at one and only one point between the two horizons $u_{ch}$ and $u_{eh}$ which provides the point at which the potential $V_\text{eff}$ has a local maximum and is, by this fact, the location of the photon sphere $u_{ps}$.

Therefore there is no stable closed orbit for the photons. If $E^2=V_\text{eff max}$, the photons describe unstable circular orbits. If $E^2<V_\text{eff max}$, the motion will be confined between the event horizon and the smaller root of $V_\text{eff}=E^2$ or between the cosmological horizon and the larger root of $V_\text{eff}=E^2$. If $E^2>V_\text{eff max}$, the motion will be confined between the event and cosmological horizons.

\begin{figure}[!htb]
    \centering
    \includegraphics[width=0.55\textwidth]{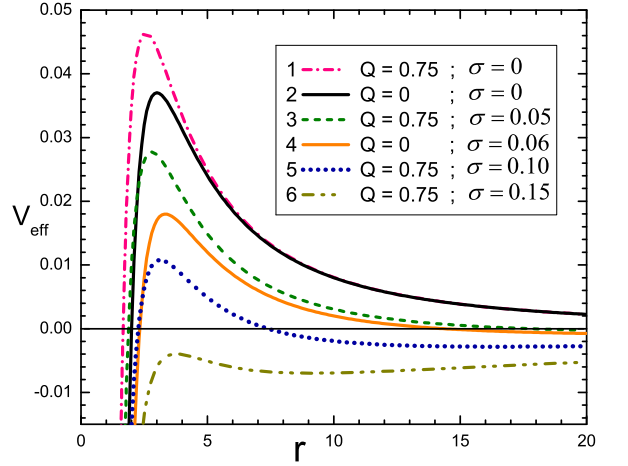}
    \caption{Effective potential ${V}_\text{eff}$ is shown as a
        function of distance $r$ for non-extreme case at different values of quintessence parameter $\sigma$ with a fixed value of the charge $Q$. $1^\text{st}$ upper curve for $V^\text{R}_\text{eff}$~, $2^\text{nd}$ curve for $V^\text{S}_\text{eff}$ and $4^\text{th}$ curve for $V^\text{K}_\text{eff}$ are taken as a reference. $3^\text{rd}$, $5^\text{th}$ and $6^\text{th}$ curves are the nonextreme case of charged Kiselev black hole effective potential $V_\text{eff}$.}
    \label{eff1}
 \end{figure}

\begin{figure}[!htb]
    \centering
    \includegraphics[width=0.55\textwidth]{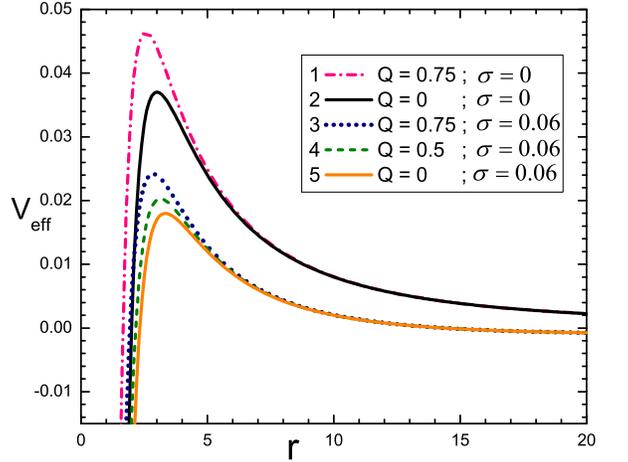}
    \caption{Effective potential ${V}_\text{eff}$ is shown as a
        function of distance $r$ for non-extreme case at a different values of charge $Q$ with constant value of quintessence parameter $\sigma$. $1^\text{st}$ upper curve for $V^\text{R}_\text{eff}$~, $2^\text{nd}$ curve for $V^\text{S}_\text{eff}$
        and $5^\text{th}$ curve for $V^\text{K}_\text{eff}$ are taken as a reference.   $3^\text{rd}$ and $4^\text{th}$ curves are the nonextreme case of charged Kiselev black hole effective potential $V_{eff}$.}
    \label{eff2}
 \end{figure}

In~\eqref{e1}, if we take $Q=0$, the effective potential reduces to the KBH effective potential,
\begin{eqnarray}\label{e2}
V^\text{K}_\text{eff}&=&\frac{L^{2}}{r^{2}}\Big(1-\frac{2M}{r}-\frac{\sigma}{r^{3w_{q}+1}}\Big).
\end{eqnarray}
When we take $\sigma=0$, (\ref{e1}) reduces to the Reissner-Nordstr\"om BH effective potential for photons,
\begin{eqnarray}\label{e33}
V^\text{R}_\text{eff}&=&\frac{L^{2}}{r^{2}}\Big(1-\frac{2M}{r}+\frac{Q^{2}}{r^{2}}\Big).
\end{eqnarray}
Further, when $\sigma=Q=0$, the effective potential for the Schwarzschild BH is given by
\begin{eqnarray}\label{e44}
 V^\text{S}_\text{eff}&=&\frac{L^{2}}{r^{2}}\Big(1-\frac{2M}{r}\Big).
\end{eqnarray}
In Figs.~\ref{eff1} and \ref{eff2},
    the effective potential $V_\text{eff}$, i.e. Eq. (\ref{e1}), is plotted
    to study the behaviour of photons near a charged KBH for the non-extreme case where $0<\sigma<0.17$ and $0<Q<1$. We observe that in each curve, there are no minima. In these graphs each curve corresponds to the maximum value $V_\text{max}$ which means that for photons, only an unstable circular orbit exists. In these two figures, the effective potentials of Kiselev (\ref{e2}), Reissner-Nordstr\"{o}m (\ref{e33}) and Schwartzschild  (\ref{e44}) black holes are displayed as references. In Fig. $\ref{eff1}$ (Fig.~$\ref{eff2}$), the quintessence parameter $\sigma$ is varying (fixed) and the charge $Q$ is fixed (varying). Both graphs are reciprocal to each other. We observe that by increasing the value of  $\sigma$ ($Q$), the photon has more (less) possibility to fell into the black hole.


\subsection{The $\pmb{u}$-$\pmb{\phi}$ trajectory equation}
In terms of $u=1/r$, we rewrite~\eqref{rdot} as
\begin{equation}\label{e5}
    \dot{u}^2=L^2u^4\Big(\frac{1}{b^2}-u^2f\Big).
\end{equation}
Combining this with~\eqref{tp2} we obtain the following equations:
\begin{align}
\Big(\frac{du}{d\phi}\Big)^2=&\ \frac{1}{b^2}-u^2f,\nn\\
\label{e6a}\quad =&\ \frac{1}{b^2}+\sigma u^{3w_q+3}-u^2+2Mu^3-Q^2u^4,\\
\frac{d^2u}{d\phi^2}+u=&\ u(1-f)-\frac{u^2}{2}\frac{df}{du},\nn\\
\label{e6b}\quad =&\ \frac{3(w_q+1)\si}{2}~u^{3w_q+2}+3Mu^2-2Q^2u^3.
\end{align}

\section{Solution to the trajectory equation}

The presence of a cosmological horizon does not make sense to investigate the photon paths beyond it as is the case beyond the event horizon. The usual bending formula~\cite{weinb}, developed for asymptotically flat solutions, no longer applies. The bending angle may be derived upon integrating either~\eqref{e6a} or~\eqref{e6b}. The usually used approach is that of Ishak and Rindler~\cite{IB}. In the so-far treated problems, $\si$ is zero, so the approach consists in integrating
\begin{equation*}
    \frac{d^2u}{d\phi^2}+u=0,\quad\text{with}\quad u(\phi=\pi/2)=b=R
\end{equation*}
which yields $u_0=\sin\phi/R$, then construct by a perturbation approach a solution to
\begin{equation}\label{ba1}
    \frac{d^2u}{d\phi^2}+u=3Mu^2-2Q^2u^3,
\end{equation}
of the form $u=u_0+u_1$, where in~\eqref{ba1} the terms in $u^n$ with $n>1$ are seen as perturbations in the limit $u\to 0$.

The approach described above does not hold if $\si\neq 0$ and $-1\leq w_q<-1/3$; since $-1\leq 3w_q+2<1$, the term proportional to $\si$ in~\eqref{e6b} is rather a leading term in the limit $u\to 0$. In the presence of quintessence, one should first solve
\begin{equation}\label{ba2a}
    \frac{d^2u_0}{d\phi^2}+u_0=\frac{3(w_q+1)\si}{2}~u_0^{3w_q+2},
\end{equation}
or
\begin{equation}\label{ba2b}
    \frac{d^2u_0}{d\phi^2}=\frac{3(w_q+1)\si}{2}~u_0^{3w_q+2},
\end{equation}
(with $-1\leq w_q<-1/3$) as if $M=0$ and $Q=0$, then by a perturbation approach one solves~\eqref{e6b}. Unfortunately, Eqs.~\eqref{ba2a} and~\eqref{ba2b} are not tractable analytically except in the cases $w_q=-1$ and $w_q=-2/3$.

In the tractable case $w_q=-2/3$, Eq.~\eqref{ba2a} reduces to
\begin{equation}\label{ba3a}
    \frac{d^2u_0}{d\phi^2}+u_0-\frac{\si}{2}=0,
\end{equation}
and it possesses the particular solution
\begin{equation}\label{ba3b}
    u_0=c \sin\phi +\frac{\si}{2}.
\end{equation}
Since~\eqref{e6b} is not equivalent to~\eqref{e6a}, from which it has been derived, one determines $c$ from the reduced expression of~\eqref{e6a} upon taking $M=0$ and $Q=0$:
\begin{equation}\label{ba3c}
    \Big(\frac{du_0}{d\phi}\Big)^2=\frac{1}{b^2}+\sigma u_0-u_0^2.
\end{equation}
Substituting~\eqref{ba3b} into~\eqref{ba3c}, we obtain
\begin{equation}\label{ba4a}
    c=\frac{\sqrt{1+\tfrac{b^2\si^2}{4}}}{b},
\end{equation}
and
\begin{equation}\label{ba4b}
    u_0=\frac{\sqrt{1+\tfrac{b^2\si^2}{4}}}{b}~\sin\phi +\frac{\si}{2}.
\end{equation}

\begin{figure}[!htb]
\centering
\includegraphics[width=0.45\textwidth]{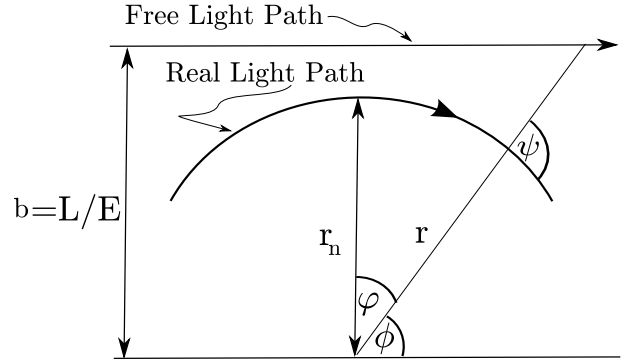}\\
\caption{\footnotesize{The diagram shows the real curved path the photons follow versus the fictitious free straight path they would follow in empty space if carrying the same angular momentum $L$ and same energy $E$. Here $\phi+\vp=\pi/2$ and $\tan\psi =r \sqrt{f} \,|d\phi/d r|=u \sqrt{f} \,|d\phi/d u|$~\eqref{psi}.}}\label{Fig2}
\end{figure}

Figure~\ref{Fig2} shows the real path that the light follows and the path that the light would follow in empty space ($\si=0,M=0,Q=0$) for the same value of the physical ratio $b=L/E$ of the angular momentum and energy. From that figure, it is obvious that in empty space $c=1/b$, which is the same expression Eq.~\eqref{ba4a} reduces to on setting $\si=0$.

If quintessence is the unique acting force ($M=0$ and $Q=0$), the minimum distance of approach $r_n$, as shown in Fig.~\ref{Fig2}, corresponds to $\phi=\pi/2$ and is derived from~\eqref{ba4b} by
\begin{equation}\label{near}
    r_n=\frac{2}{2c+\si}=\frac{b}{\sqrt{1+\tfrac{b^2\si^2}{4}}+\tfrac{b\si}{2}}<b.
\end{equation}
This can be derived directly from the definition of $r_n$, which is the nearest distance from the light path to the lens. This is such that the r.h.s of~\eqref{e6a} is 0, yielding the same expression as in~\eqref{near}.

In bending-angle problems the parameter $b$ is assumed to be large
to allow for series expansions in powers of $1/b$. Since
quintessence is not supported observationally, we make the statement
that $\si\ll 1$, which we will make clearer in the next section
[Eq.~\eqref{m6}].

All authors who worked on the bending angle in a de Sitter-like
geometry draw similar figure as Fig.~\ref{Fig2}~, but they make no
distinction between $b$ and $r_{n}$; rather, they use loosely a
common notation $R$ for $b$ and $r_{n}$. This remains more or less
justified as far as quintessence is not taken into consideration
where one may write $b \gtrsim r_{n}$. As we mentioned earlier, in
bending-angle problems the parameter $b$ is assumed large to allow
for series expansions in powers of $1/b$, so in presence of
quintessence, one has to further assume $b\si=\tfrac{L\si}{E}\ll 1$
[Eq.~\eqref{m6}] in order to have $b \gtrsim r_{n}$. In presence of
quintessence, corrections in the expression of $r_n$ are needed: If
$\si\ll 1$ and $b\si\ll 1$ we obtain to the first order in $1/b$
[see Eq.~\eqref{ba8cun} for further orders of approximation]
\begin{equation}\label{ba5b}
u_n=\frac{1}{r_n}= \frac{1}{b}\big[1+\frac{b\si}{2}+\frac{b^2\si^2}{8}+\mathcal{O}(b^4\si^4)\big].
\end{equation}

Now, substituting $u=u_0+u_1$ into~\eqref{e6b} reduces to ($w_q=-2/3$)
\begin{equation*}
    \frac{d^2u_1}{d\phi^2}+u_1=3Mu_0^2-2Q^2u_0^3,
\end{equation*}
where $u_0$ is given by~\eqref{ba4b}. A particular exact solution is
\begin{equation}\label{ba6}
u_1=\frac{3M\si^2}{4}-\frac{Q^2\si^3}{4}+cC_1+c^2C_2+c^3C_3,
\end{equation}
where $c$ is given by~\eqref{ba4a} and the coefficients ($C_1,C_2,C_3$) are related to the coefficients ($B_1,B_2,B_3$), which were first evaluated in Ref.~\cite{FMR}, by $C_1=B_1-\sin\phi$, $C_2=B_2$, and $C_3=B_3$.
\begin{align}
&C_1 =   M \big( \tfrac{ 3 \sigma \pi}{4} \cos \phi - \tfrac{3}{2} \phi \sigma \cos \phi + \tfrac{3}{2} \sigma \sin \phi \big)\nn\\
&\qquad\ +Q^2\big( \tfrac{3}{4} \phi \sigma^2 \cos \phi - \tfrac{3}{8} \pi \sigma^2 \cos \phi  - \tfrac{3}{4} \sigma^2 \sin \phi \big),\nn\\
\label{ba7}&C_2 = M \big( \tfrac{3}{2} + \tfrac{1}{2} \cos 2 \phi \big) - Q^2 \big( \tfrac{ 3 \sigma}{2} + \tfrac{ \sigma}{2} \cos 2 \phi \big),\\
&C_3 = Q^2 \big( \tfrac{3}{4} \phi \cos \phi- \tfrac{ 3}{8} \pi \cos \phi  -\tfrac{9}{16} \sin \phi - \tfrac{1}{16} \sin 3 \phi \big).\nn
\end{align}

Under the constraints $\si\ll 1$ and $b\si\ll 1$, expansions of the r.h.s of~\eqref{ba7} and of $u_0$~\eqref{ba4b} yield
\begin{align}\label{ba8c}
&u= \tfrac{1}{b} [\sin\phi+\tfrac{1}{2}b\si+\tfrac{\sin\phi}{8} b^2\si^2+\mathcal{O}(b^4 \sigma ^4)]\nn\\
&+\tfrac{M}{b^2} [\tfrac{3}{2}+\tfrac{1}{2} \cos 2\phi+(\tfrac{3}{4} \pi  \cos\phi-\tfrac{3}{2} \phi  \cos\phi+\tfrac{3}{2}
\sin\phi) b\si\nn\\
&+(\tfrac{9}{8}+\tfrac{1}{8} \cos 2\phi) b^2\si^2+\mathcal{O}(b^3 \sigma ^3)]\\
&+\tfrac{Q^2}{b^3}[\tfrac{3}{4} \phi  \cos\phi-\tfrac{3}{8} \pi  \cos\phi-\tfrac{9}{16} \sin\phi-\tfrac{1}{16} \sin 3\phi\nn\\
&-(\tfrac{3}{2}+\tfrac{1}{2} \cos 2\phi) b\si+\mathcal{O}(b^2 \sigma ^2)].\nn
\end{align}
\begin{equation}
u_0\simeq \tfrac{1}{b} [\sin\phi+\tfrac{1}{2}b\si+\tfrac{\sin\phi}{8} b^2\si^2].
\end{equation}
For~\eqref{ba8c} to hold it is sufficient that the products $M\si\,b\si$ and $Q^2\si^2\,b\si$ remain much smaller than unity. This conclusion is easily derived from the requirement that $u_1/u_0\ll 1$.

As to the minimum distance $r_n=1/u_n$, this is given by [setting $\phi=\pi/2$ in~\eqref{ba8c}]
\begin{align}\label{ba8cun}
u_n & =  \tfrac{1}{b} [1+\tfrac{1}{2}b\si+\tfrac{1}{8} b^2\si^2+\mathcal{O}(b^4 \sigma ^4)]\nn\\
&+\tfrac{M}{b^2} [1+\tfrac{3}{2} b\si +b^2\si^2+\mathcal{O}(b^3 \sigma ^3)]\\
&-\tfrac{Q^2}{b^3}[\tfrac{1}{2}+ b\si+\mathcal{O}(b^2 \sigma ^2)].\nn
\end{align}
We see that the mass $M$ contributes to the second order while $Q^2$ contributes to the third order of the series expansion in powers of $1/b$.

\begin{figure*}[!htb]
\centering
\includegraphics[width=0.65\textwidth]{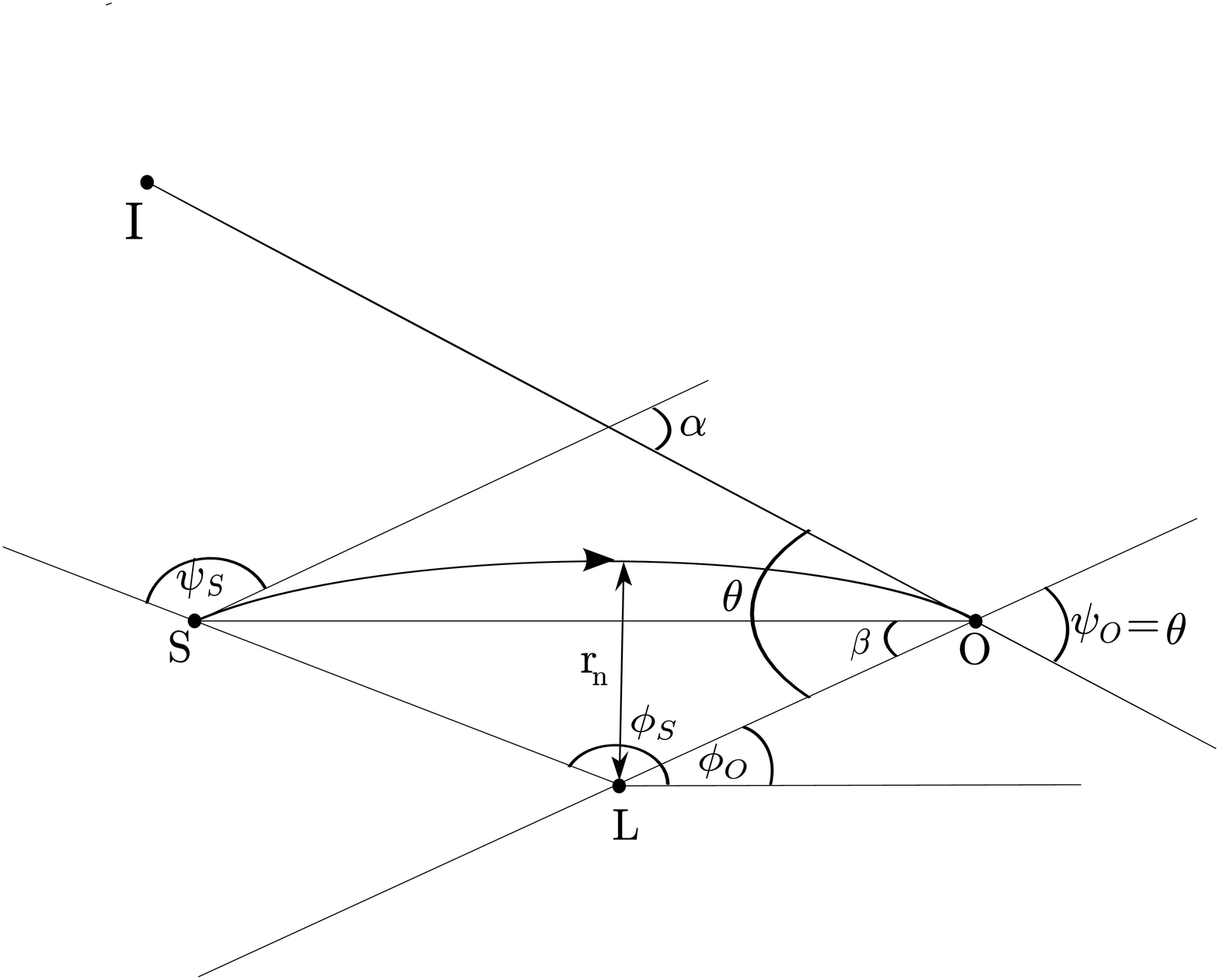}\\
\caption{\footnotesize{The symbols I, L, O, and S denote the image, lens (black hole), observer, and source, respectively. The angles $\phi$ and $\psi$, $r_{\text{min}}$, and $r$ are those defined in Fig.~\ref{Fig2}. The angles $\bt$ and $\ta$ are the angular positions of the source and image, and $\al=(\phi_s-\phi_o)+(\psi_o-\psi_s)$ is the deflection angle. The image location $\ta$ is the angle $\angle\,\text{IOL}$, which is by definition $\psi_o=\ta$.}}\label{Fig3}
\end{figure*}

\section{Lens equation - Bending angle}

The expression of the angle $\psi$ defined as the angle the direction $\phi$ makes with the light path at $r$, as depicted in Fig.~\ref{Fig2}, is given by~\cite{R}
\begin{equation*}
    \tan\psi =r \sqrt{f} \Big|\frac{d\phi}{d r}\Big|=u \sqrt{f} \Big|\frac{d\phi}{d u}\Big|,
\end{equation*}
or, preferably, by~\cite{GB}
\begin{equation}\label{psi}
    \sin\psi = bu\sqrt{f(u)}.
\end{equation}
Series expansion of $\phi$ may be determined upon reversing the expansion~\eqref{ba8c}. This is a cumbersome work which we will avoid in this section. Rather, we will rely on~\eqref{psi} and on the integral form of $\phi$~\eqref{e6a},
\begin{equation}
\phi=\int\frac{du}{\sqrt{\frac{1}{b^2}-u^2f}},
\end{equation}
to determine the deflection angle $\al$.

Figure~\ref{Fig3} depicts a light path along with the locations of the lens (L: black hole), observer (O), source (S), and image (I). The observer sees the image along the direction OI, which is tangent to the light path at O. The angles $\bt$ and $\ta$ are the angular positions of the source and image. The image location $\ta$ is the angle $\angle\,\text{IOL}$, which is by definition $\psi_o=\ta$. The distances from the lens to the observer and to the source are denoted by $r_o=1/u_o$ and $r_s=1/u_s$, respectively. The nearest distance from the light path to the lens, denoted by $r_n=1/u_n$, is such that the r.h.s of~\eqref{e6a} is 0, yielding
\begin{equation}\label{m1}
\frac{1}{b^2}=u_n^2f(u_n).
\end{equation}
In the special case $w_{q}=-2/3$, we obtain $1/b^2=u_n(u_n-2Mu_n^2+Q^2u_n^3-\sigma)$ a series solution of which is given by~\eqref{ba8cun}. In this section, instead of $b$, we will employ $u_n$ as an independent parameter around which we expand the deflection angle $\al$.

Let $F(u)$ denote the function on the r.h.s of~\eqref{e6a}
\begin{equation}\label{m2}
F(u)=u_n^2f(u_n)-u^2f(u),
\end{equation}
where we have used~\eqref{m1}. From Fig.~\ref{Fig3}, we see that the deflection angle $\al$ is given by
\begin{equation}\label{m3}
    \al=\int_{u_o}^{u_n}\frac{\dd u}{\sqrt{F}} +\int_{u_s}^{u_n}\frac{\dd u}{\sqrt{F}}+\psi_o-\psi_s,
\end{equation}
where the sum of the first two terms is, according to Fig.~\ref{Fig3}, is the integral form of the angle $\angle\,\text{SLO}=\phi_s-\phi_o$. Equivalently, Eq.~\eqref{m3} is brought to the form
\begin{equation}\label{m4}
    \al=2\int_{0}^{u_n}\frac{\dd u}{\sqrt{F}} - \int_{0}^{u_o}\frac{\dd u}{\sqrt{F}} - \int_{0}^{u_s}\frac{\dd u}{\sqrt{F}}+\psi_o-\psi_s.
\end{equation}
By Fig.~\ref{Fig3} and Eq.~\eqref{psi} we have $\psi_o=\arcsin (bu_o\sqrt{f(u_o)})$ and $\psi_s=\pi-\arcsin (bu_s\sqrt{f(u_s)})$. Using~\eqref{m1} in these expressions we arrive at
\begin{align}\label{m5}
&\al=2\int_{0}^{u_n}\frac{\dd u}{\sqrt{u_n^2f(u_n)-u^2f(u)}} -\pi\nn\\
&- \int_{0}^{u_o}\frac{\dd u}{\sqrt{u_n^2f(u_n)-u^2f(u)}}- \int_{0}^{u_s}\frac{\dd u}{\sqrt{u_n^2f(u_n)-u^2f(u)}}\nn\\
&+\arcsin \Big(\frac{u_o}{u_n}\sqrt{\frac{f(u_o)}{f(u_n)}}\Big)
+\arcsin \Big(\frac{u_s}{u_n}\sqrt{\frac{f(u_s)}{f(u_n)}}\Big).
\end{align}
The first line in~\eqref{m5} is the expression of the deflection angle we would have obtained had we assumed the observer and the source to be at spatial infinity ($u_o\equiv 0$ and $u_s\equiv 0$). The four last terms in~\eqref{m5} are corrections added to the asymptotically-flat expression of the deflection angle. From now on, we assume that the \emph{independent} parameters ($u_o\ll 1,u_s\ll 1,u_n\ll 1$) are small compared to unity but are not 0.

Another important parameter is $\si$. Since quintessence has not been observed in the cosmos, it is legitimate to assume $\si\ll 1$; rather, we assume
\begin{equation}\label{m6}
   \si\ll \min (u_o,u_s)<\max (u_o,u_s)\ll u_n\ll 1,
\end{equation}
considering thus quintessence as a perturbation to the Reissner-Nordstr\"om black hole (the constarint $\max (u_o,u_s)\ll u_n$ is satisfied by definition of $u_n$). The evaluation of~\eqref{m5} consists in determining the series expansion of its r.h.s in powers of the \emph{independent} parameters ($\si\ll 1,u_o\ll 1,u_s\ll 1,u_n\ll 1$).

We will not assume the location of the observer to correspond to $\phi_o=\pi/2$, as some authors did~\cite{RI,FMR}, for this introduces a wrong term~\cite{BBS} in the series expansion\footnote{This is similar to finding a series expansion to the third order in $x$ of, say, $\ln(1+\sin x)$. Expanding $\ln(1+\sin x)$ as $\sin x-\sin^2 x/2$ produces the wrong answer: $\ln(1+\sin x)\simeq x -x^2/2-x^3/6$. The correct step is to expand $\ln(1+\sin x)$ by $\ln(1+\sin x)\simeq \sin x-\frac{\sin^2 x}{2}+\frac{\sin^3 x}{3}$,\
which yields $ \ln(1+\sin x)\simeq x-\frac{x^2}{2}+\frac{x^3}{6}$.}
%
of $\al$.

\section{Strong deflection limit}
\subsection{Case $w_{q}=-2/3$}\label{wq23}
Conditions~\eqref{m6} being observed, we find in the case $w_{q}=-2/3$:
\begin{align}\label{s1}
\alpha & =  4 M u_n+[(15 \pi -16)M^2 -3 \pi  Q^2] \frac{u_n^2}{4}\nn\\
\; &-\frac{M (1+M u_n) (u_o^2+u_s^2)}{u_n}\\
\; &+\Big\{\frac{(3 \pi -4)M}{2}+[(88-15 \pi )M^2 -(16-3 \pi ) Q^2]\frac{u_n}{4} \nn\\
\; &-\frac{M^2 (u_o+u_s)}{2}-\frac{2 M  (u_o^2+u_s^2)}{4u_n^2}-\frac{M (u_o+u_s)}{2 u_n}\Big\} \sigma.\nn
\end{align}
In the first line we recognize the expression of the deflection angle for the Reissner-Nordstr\"om black hole as determined in Ref.~\cite{paths} (in Eq.~(2.8) of Ref.~\cite{paths}, $\ga=1$ corresponds to Reissner-Nordstr\"om black hole and to obtain the first line in~\eqref{s1} from Eq.~(2.8) of Ref.~\cite{paths} insert $r_++r_-=2M$, $r_+r_-=Q^2$, and $r_+^2+r_-^2=4M^2-2Q^2$, where $r_-<r_+$ are the two horizons). The second line in~\eqref{s1} is a correction to the deflection angle for the Reissner-Nordstr\"om black hole when the observer and the source are at large, but finite, distances from the lens. Notice that this correction up to the power 2 in $u_o$ and $u_s$ does not depend on the charge of the black hole. The remaining terms, proportional to $\si$, are corrections due to quintessence.

The power series in the r.h.s of~\eqref{s1} has been determined as follows. The series expansions of the arcsin terms in~\eqref{m5} in powers of ($\sigma,u_o,u_s,u_n$) is straightforward; the series expansion of the first line in~\eqref{m5} has been done in Appendix A of Ref.~\cite{paths}. In this work we show how to derive the series expansion of the first term in the second line of~\eqref{m5}; the series expansion of the second term is obtained by mere substitution $u_o\leftrightarrow u_s$. In all calculations the series expansions are obtained in the order given in~\eqref{m6}; that is, we first expand with respect to $\si$ to order 1, then expand with respect to ($u_o,u_s$) to order 2, and finally we expand with respect to $u_n$ to order 2 too. In the final expansion~\eqref{s1} we have kept all the terms with order not exceeding 2.

Set $u=u_ox$ and $0\leq x\leq 1$. The first term in the second line of~\eqref{m5} becomes
\begin{equation}\label{s2}
    u_o\int_{0}^{1}\frac{\dd x}{\sqrt{u_n^2f(u_n)-u_o^2x^2f(u_ox)}},
\end{equation}
yielding in the case $w_{q}=-2/3$
\begin{multline}\label{s3}
\frac{u_o}{\sqrt{u_n^2f(u_n)-u_o^2x^2f(u_ox)}}\simeq \Big(M+\frac{15 M^2 \sigma }{4}-\frac{3 Q^2 \sigma }{4}\Big)u_o\\
+\frac{u_o \sigma}{2 u_n^2}+ (3 M^2-Q^2)~\frac{u_n u_o}{2}+\frac{(2+3 M \sigma )u_o}{2 u_n},
\end{multline}
where the integration over $x$ is straightforward.

\subsection{Case for all $-1\leq w_{q}< -1/3$}
In this section, we will obtain the general expression for the deflection angle for any value of $w_{q}$. For simplicity, let us introduce a new constant
\begin{eqnarray}
\gamma=3(w_{q}+1)\,,
\end{eqnarray}
which makes Eq. (\ref{e6a}) easier to handle. Now, all the exponents in $u$ in (\ref{e6a}) are positive or zero. Since $-1\leq w_{q}< -1/3$, the new constant lies between $0\leq\gamma< 2$.
Now, let us compute each term of Eq. (\ref{m5}) separately. For sake of simplicity we will name each term of (\ref{m5}) as follows,
\begin{align}
&I_{1}=\int_{0}^{u_n}\frac{\dd u}{\sqrt{u_n^2f(u_n)-u^2f(u)}}\,,\label{I1}\\
&I_{2}=- \int_{0}^{u_o}\frac{\dd u}{\sqrt{u_n^2f(u_n)-u^2f(u)}}\nonumber\\
&\quad - \int_{0}^{u_s}\frac{\dd u}{\sqrt{u_n^2f(u_n)-u^2f(u)}}\,,\label{I2}\\
&I_{3}=\arcsin \Big(\frac{u_o}{u_n}\sqrt{\frac{f(u_o)}{f(u_n)}}\Big)
+\arcsin \Big(\frac{u_s}{u_n}\sqrt{\frac{f(u_s)}{f(u_n)}}\Big)\,,\label{I3}
\end{align}
so that Eq. (\ref{m5}) can be expressed as
\begin{eqnarray}
\alpha=2I_{1}-\pi+I_{2}+I_{3}\,.\label{alphaa}
\end{eqnarray}
The final expression for $\alpha$ needs to be separated in three ranges of $\gamma$: i) $\gamma=0$ , ii) $0<\gamma\leq 1$ and iii) $1<\gamma<2$. In the following sections, we will follow the same idea as in Sec.~\ref{wq23} to compute all these terms for any $\gamma$.

\subsubsection{ Computing $I_{1}$}
Let $u=u_{n}x$ with $0\leq x \leq 1$. First, we expand the integrand of $I_{1}$ up to first order in $\sigma$ and then up to second order in $u_{n}$. By doing that, for $0\leq\gamma\leq 1$ the expansion of the integrand of (\ref{I1}) takes the form
\begin{widetext}
\begin{eqnarray}
\frac{u_n}{\sqrt{u_n^2f(u_n)-u_n^2x^2f(u_nx)}}&\simeq& \frac{3 \sigma  u_{n}^{\gamma } \left(x^{\gamma }-1\right) \left(Q^2 (x+1)^2 \left(x^2+1\right)-5 M^2 \left(x^2+x+1\right)^2\right)}{4 (1-x)^{3/2} (x+1)^{7/2}}\nonumber\\
&&+\frac{u_{n}^2 \left(3 M^2 \left(x^2+x+1\right)^2-Q^2 (x+1)^2 \left(x^2+1\right)\right)}{2 \sqrt{1-x} (x+1)^{5/2}}+\frac{1}{\sqrt{1-x^2}}\nonumber\\
&&-\frac{3 M \sigma  \left(x^2+x+1\right) u_{n}^{\gamma -1} \left(x^{\gamma }-1\right)}{2 (1-x)^{3/2} (x+1)^{5/2}}+\frac{M u_{n} \left(x^2+x+1\right)}{\sqrt{1-x} (x+1)^{3/2}}-\frac{\sigma  u_{n}^{\gamma -2} \left(x^{\gamma }-1\right)}{2 \left(1-x^2\right)^{3/2}}\,,
\end{eqnarray}
\end{widetext}
and for $1<\gamma<2$,
\begin{widetext}
\begin{eqnarray}
\frac{u_n}{\sqrt{u_n^2f(u_n)-u_n^2x^2f(u_nx)}}&\simeq&\frac{u_{n}^2 \left(3 M^2 \left(x^2+x+1\right)^2-Q^2 (x+1)^2 \left(x^2+1\right)\right)}{2 \sqrt{1-x} (x+1)^{5/2}}+\frac{1}{\sqrt{1-x^2}}
\nonumber\\
&&-\frac{3 M \sigma  \left(x^2+x+1\right) u_{n}^{\gamma -1} \left(x^{\gamma }-1\right)}{2 (1-x)^{3/2} (x+1)^{5/2}}
+\frac{M u_{n} \left(x^2+x+1\right)}{\sqrt{1-x} (x+1)^{3/2}}+\frac{\sigma  u_{n}^{\gamma -2}(1-x^{\gamma})}{2 \left(1-x^2\right)^{3/2}}\,.
\end{eqnarray}
\end{widetext}
Integration over $x$ will depend on $\gamma$ so that it is not possible to write down an explicit result for $I_{1}$ for a general $\gamma$. Therefore, for $0\leq\gamma\leq 1$, we can write $I_{1}$ as follows
\begin{widetext}
\begin{eqnarray}\label{I1fa}
I_{1}&\simeq&-\int_0^1 \frac{3 M \sigma  \left(x^2+x+1\right) u_{n}^{\gamma -1} \left(x^{\gamma }-1\right) \left(5 M u_{n} \left(x^2+x+1\right)+2 (x+1)\right)}{4 (1-x)^{3/2} (x+1)^{7/2}} \, dx+u_{n}^2 \left(\left(\frac{15 \pi }{8}-2\right) M^2-\frac{3 \pi  Q^2}{8}\right)\nonumber\\
&&+2 M u_{n}+\frac{3}{4} Q^2 \sigma  \left(\frac{\pi }{2}-\frac{\sqrt{\pi } (2 \gamma +1) \Gamma \left(\frac{\gamma +1}{2}\right)}{\gamma  \Gamma \left(\frac{\gamma }{2}\right)}\right) u_{n}^{\gamma }+\frac{\sqrt{\pi } \sigma  \,\Gamma \left(\frac{\gamma +1}{2}\right) u_{n}^{\gamma -2}}{2 \,\Gamma \left(\frac{\gamma }{2}\right)}+\frac{\pi }{2}\,.
\end{eqnarray}
\end{widetext}
Note that $\displaystyle\lim_{\gamma\to 0}\Gamma((\gamma+1)/2)/\Gamma(\gamma/2)=0$ and $\displaystyle\lim_{\gamma\to 0}\Gamma((\gamma+1)/2)/(\gamma\Gamma(\gamma/2))=\sqrt{\pi}/2$ are finite, so that the above expression is well defined for $\gamma=0$. Now, for the range $1<\gamma< 2$, the integral becomes
\begin{widetext}
\begin{eqnarray}\label{I1fb}
I_{1}&\simeq&   \int_0^1 -\frac{3 M \sigma  \left(x^2+x+1\right) u_{n}^{\gamma -1} \left(x^{\gamma }-1\right)}{2 (1-x)^{3/2} (x+1)^{5/2}} \, dx+u_{n}^2 \left(\left(\frac{15 \pi }{8}-2\right) M^2-\frac{3 \pi  Q^2}{8}\right)+2 M u_{n}\nonumber\\
&&+\frac{\sqrt{\pi } \sigma \, \Gamma \left(\frac{\gamma +1}{2}\right) u_{n}^{\gamma -2}}{2 \, \Gamma \left(\frac{\gamma }{2}\right)}+\frac{\pi }{2}\,.
\end{eqnarray}
\end{widetext}

\subsubsection{ Computing $I_{2}$}
First, we will compute the first term of $I_{2}$ and then we can directly use that result to compute the second term of $I_{2}$ by changing $u_{o}$ for $u_{s}$. As we did before, we set $u=u_o x $ and expand up to first order in $\sigma$ and then up to second order in $u_{0}$. Finally, we need to take expansions up to second order in $u_{n}$. The integrand of the first term of $I_{2}$ is then expanded as follows,
\begin{widetext}
\begin{multline}
\frac{u_o}{\sqrt{u_n^2f(u_n)-u_o^2x^2f(u_ox)}}\simeq
\begin{cases}
\frac{5}{4} M \sigma  u_{o} \left(7 M^2-3 Q^2\right)+\frac{1}{4} M u_{o} \left(-35 M^2 \sigma +15 Q^2 \sigma +4\right)\\
+\frac{3 \sigma  u_{o} \left(5 M^2-Q^2\right)}{4 u_{n}}+\frac{u_{o} \left(-15 M^2 \sigma +3 Q^2 \sigma +4\right)}{4 u_{n}}-\frac{1}{2} u_{o} u_{n} \left(Q^2-3 M^2\right)\,, & \gamma=0\\ \\
\frac{u_{o} \left(-15 M^2 \sigma  u_{o}^{\gamma } x^{\gamma }+3 Q^2 \sigma  u_{o}^{\gamma } x^{\gamma }+4\right)}{4 u_{n}}+\frac{3}{4} \sigma  u_{o} \left(5 M^2-Q^2\right) u_{n}^{\gamma -1}+M u_{o}\\
-\frac{1}{2} u_{o} u_{n} \left(Q^2-3 M^2\right)
+\frac{1}{2} \sigma  u_{o} (3 M u_{n}+1) u_{n}^{\gamma -3}\,,& 0<\gamma < 1\\ \\
\big(M+\frac{15 M^2 \sigma }{4}-\frac{3 Q^2 \sigma }{4}\big)u_o
+\frac{u_o \sigma}{2 u_n^2}+ (3 M^2-Q^2)~\frac{u_n u_o}{2}+\frac{(2+3 M \sigma )u_o}{2 u_n}\,,& \gamma = 1\\ \\
-\frac{1}{2} u_{o} u_{n} \left(Q^2-3 M^2\right)
+\frac{1}{2} \sigma  u_{o} (3 M u_{n}+1) u_{n}^{\gamma -3}+M u_{o}+\frac{u_{o}}{u_{n}}\,, & 1<\gamma<2
\end{cases}\,.
\end{multline}
\end{widetext}
Therefore, by integrating over $x$ and then compute the second integral by changing $u_{o}$ by $u_{s}$ we arrive at
\begin{widetext}
    \begin{multline}
I_{2}\simeq\begin{cases}
-\frac{1}{2} u_{n} \left(3 M^2-Q^2\right) (u_{o}+u_{s})-M (u_{o}+u_{s})-\frac{u_{o}+u_{s}}{u_{n}}\,, & \gamma=0\\ \\
\frac{3
    (u_{0}^{1+\gamma}+u_{s}^{1+\gamma})\sigma \left( 5 M^2-Q^2  \right)}{4 (\gamma +1) u_{n}}-\frac{3}{4} \sigma  \left(5 M^2-Q^2\right) (u_{o}+u_{s}) u_{n}^{\gamma -1}\\
-\frac{1}{2} u_{n} \left(3 M^2-Q^2\right) (u_{o}+u_{s})+\frac{3 M \sigma  (u_{o}^{\gamma +1}+u_{s}^{\gamma +1})
}{2 (\gamma +1) u_{n}^2}-\frac{3}{2} M \sigma  (u_{o}+u_{s}) u_{n}^{\gamma -2}-M (u_{o}+u_{s})\\
+\frac{\sigma  \left(u_{o}^{\gamma +1}+u_{s}^{\gamma +1}\right)}{2 (\gamma +1) u_{n}^3}-\frac{1}{2} \sigma  (u_{o}+u_{s}) u_{n}^{\gamma -3}-\frac{u_{o}+u_{s}}{u_{n}}\,,& 0<\gamma<1\\ \\
-\frac{1}{2} u_{n} \left(3 M^2-Q^2\right) (u_{o}+u_{s})-\frac{3}{4} \sigma  \left(5 M^2-Q^2\right) (u_{o}+u_{s})-\frac{3 M \sigma  (u_{o}+u_{s})}{2 u_{n}}\\
-M (u_{o}+u_{s})-\frac{\sigma  (u_{o}+u_{s})}{2 u_{n}^2}-\frac{u_{s}+u_{0}}{u_n}\,,& \gamma= 1\\ \\
-\frac{1}{2} u_{n} \left(3 M^2-Q^2\right) (u_{o}+u_{s})
    -\frac{3}{2} M \sigma  (u_{o}+u_{s}) u_{n}^{\gamma -2}-M (u_{o}+u_{s})\\
    -\frac{1}{2} \sigma  (u_{o}+u_{s}) u_{n}^{\gamma -3}-\frac{u_{o}+u_{s}}{u_{n}}\,,& 1<\gamma<2
\end{cases} \,.
\end{multline}
\end{widetext}

\subsubsection{ Computing $I_{3}$}
By expanding the term $I_{3}$ as we did before, i.e., first up to first order in $\sigma$, then up to second order in $u_o$ and finally up to second order in $u_{n}$ we find
\begin{widetext}
    \begin{multline}
I_{3}\simeq\begin{cases}
-\frac{u_{n} \left(3 M^2-Q^2\right) \left(M \sigma  u_{o}+\sigma -2 u_{o}^2\right)}{4 u_{o}}+\frac{M \left(u_{o}^2 \left(29 M^2 \sigma -13 Q^2 \sigma +8\right)-8 M u_{o}^3-4 M \sigma  u_{o}-4 \sigma \right)}{8 u_{o}}\\
\frac{u_{o}^2 \left(9 M^2 \sigma -Q^2 \sigma +8\right)-8 M u_{o}^3-4 M \sigma  u_{o}-4 \sigma }{8 u_{o} u_{n}}
-\frac{3 M \sigma  u_{o} (M u_{o}-1)}{4 u_{n}^2}\,, & \gamma=0\\ \\
-\frac{3}{4} M^2 \sigma  u_{n} u_{o}^{\gamma -1}-\frac{21 M^2 \sigma  u_{o}^{\gamma +1}}{8 u_{n}}-\frac{1}{2} M^2 \sigma  u_{o}^{\gamma }-\frac{3}{2} M^2 \sigma  u_{o}^2 u_{n}^{\gamma -2}-M^2 u_{o}^2+\frac{15}{4} M^2 \sigma  u_{o} u_{n}^{\gamma -1}\\
+\frac{3}{2} M^2 u_{o} u_{n}-\frac{1}{2} M \sigma  u_{o}^{\gamma -1}-\frac{3 M \sigma  u_{o}^{\gamma +1}}{4 u_{n}^2}-\frac{M \sigma  u_{o}^{\gamma }}{2 u_{n}}-\frac{1}{2} M \sigma  u_{o}^2 u_{n}^{\gamma -3}-\frac{M u_{o}^2}{u_{n}}+\frac{3}{2} M \sigma  u_{o} u_{n}^{\gamma -2}\\
+M u_{o}+\frac{1}{4} Q^2 \sigma  u_{n} u_{o}^{\gamma -1}+\frac{5 Q^2 \sigma  u_{o}^{\gamma +1}}{8 u_{n}}-\frac{3}{4} Q^2 \sigma  u_{o} u_{n}^{\gamma -1}-\frac{1}{2} Q^2 u_{o} u_{n}
-\frac{\sigma  u_{o}^{\gamma -1}}{2 u_{n}}\\
+\frac{1}{2} \sigma  u_{o} u_{n}^{\gamma -3}+\frac{u_{o}}{u_{n}}\,,& 0<\gamma< 1\\ \\
u_{o} \left(\frac{13 M^2 \sigma }{4}+M-\frac{3 Q^2 \sigma }{4}\right)+u_{n} \left(\frac{1}{4} \sigma  \left(Q^2-3 M^2\right)+\frac{1}{2} u_{o} \left(3 M^2-Q^2\right)\right)-M^2 u_{o}^2-\frac{M \sigma }{2}\\
+\frac{1}{u_n^2}\Big(\frac{\sigma  u_{o}}{2}-\frac{1}{2} M \sigma  u_{o}^2\Big)-\frac{1}{u_n}\Big(
\frac{\sigma }{2}-u_{o} (M \sigma +1)\Big)\,,& \gamma= 1\\ \\
-M^2 u_o^2+\frac{3}{2} M^2 u_o u_n-\frac{M u_o^2}{u_n}+\frac{1}{2} \sigma  u_o u_n^{\gamma -3}+M u_o-\frac{1}{2} Q^2 u_o u_n-\frac{\sigma  u_o^{\gamma -1}}{2 u_n}+\frac{u_o}{u_n}+\frac{3}{2} M \sigma  u_{o} u_{n}^{\gamma -2}\,, & 1<\gamma<2
\end{cases}\,.
\end{multline}
\end{widetext}

\subsubsection{Computing $\alpha$}
Now, we have all the ingredients to find the final expression for $\alpha$ for a general $\gamma$. If we replace all the terms computed before in (\ref{alphaa}), for $\gamma=0$ we find \newpage
\begin{widetext}
\begin{eqnarray}\label{al1}
\alpha&=&\frac{1}{{u_{n}}}\Big[- M (u_{o}^2+u_{s}^2+\sigma) +\frac{1}{8} \sigma  (u_{o}+u_{s}) \left(9 M^2-Q^2\right)-\frac{\sigma }{2 }\Big(\frac{1}{u_{o}}+\frac{1}{u_{s}}\Big)\Big]\nonumber\\
&& +\frac{1}{8} M \sigma  (u_{s}+u_0) \left(29 M^2-13 Q^2\right)+u_{n}^2 \Big(
\big(\frac{15 \pi }{4}-4\big) M^2-\frac{3 \pi  Q^2}{4}\Big)-M^2(u_{o}^2+ u_{s}^2+\sigma)\nonumber\\
&&+u_{n} \left(\frac{1}{2} M \left((Q^2-3 M^2) \sigma  +8\right)+\frac{\sigma  \left(Q^2-3 M^2\right)}{4 }\Big(\frac{1}{u_{o}}+\frac{1}{u_{s}}\Big)\right)-\frac{M \sigma }{2 }\Big(\frac{1}{u_{o}}+\frac{1}{u_{s}}\Big) \nonumber\\
&&-\frac{3}{4u_{n}^2} M \sigma  \left(M (u_{o}^2+ u_{s}^2)-u_{s} -u_{o}\right)\,,
\end{eqnarray}
\end{widetext}
for $0<\gamma< 1$ we get
\begin{widetext}
\begin{eqnarray}
\alpha&=&-3 M \sigma u_{n}^{\gamma -1}\int_0^1 \frac{\left(x^2+x+1\right) \left(x^{\gamma }-1\right) \left(5 M u_{n} \left(x^2+x+1\right)+2 (x+1)\right)}{2 (1-x)^{3/2} (x+1)^{7/2}} \, dx \nonumber\\
&&-\frac{M \left(M \sigma  u_{o}^{\gamma}+2 M u_{o}^2+\sigma  u_{o}^{\gamma -1}\right)}{2}
-\frac{M \left(M \sigma  u_{s}^{\gamma}+2 M u_{s}^2+\sigma  u_{s}^{\gamma -1}\right)}{2}\nn\\
&&+u_{n} \left(4M+\frac{(Q^2 - 3 M^2)\sigma u_{o}^{\gamma -1}}{4}+\frac{(Q^2 - 3 M^2)\sigma u_{s}^{\gamma -1}}{4}\right)- \frac{M \sigma u_{n}^{\gamma -3}}{2}(u_{o}^2+ u_{s}^2)\nonumber\\
&&+u_{n}^2 \left(\Big(\frac{15 \pi }{4}-4\Big) M^2-\frac{3 \pi  Q^2}{4}\right)-\frac{3 (\gamma -1) M \sigma  (u_o^{\gamma +1}+u_s^{\gamma +1})}{4 (\gamma +1) u_n^2}\nonumber\\
&&
+\frac{3 \sqrt{\pi } Q^2 \sigma  \left(\sqrt{\pi } \gamma  \, \Gamma \left(\frac{\gamma }{2}\right)-2 (2 \gamma +1) \, \Gamma \left(\frac{\gamma +1}{2}\right)\right) u_n^{\gamma }}{4 \gamma  \, \Gamma \left(\frac{\gamma }{2}\right)}\nonumber\\
&&-\frac{1}{u_{n}}\Big[\frac{\sigma  [3 (7 \gamma -3) M^2+(1-5 \gamma ) Q^2] (u_o^{\gamma +1}+u_s^{\gamma +1})}{8 (\gamma +1)}+\frac{1}{2} \sigma (u_o^{\gamma -1}+u_s^{\gamma -1})\nonumber\\
&&+\frac{1}{2} M \sigma  \left(u_o^{\gamma }+u_s^{\gamma }\right)+M (u_o^2+u_s^2)\Big]+\frac{1}{2} \sigma  u_n^{\gamma -2} \Big(\frac{2 \sqrt{\pi }\, \Gamma \big(\frac{\gamma +1}{2}\big)}{\Gamma \left(\frac{\gamma }{2}\right)}-3 M^2 (u_o^2+u_s^2)\Big)\,.
\end{eqnarray}
\end{widetext}
Finally, for $1<\gamma<2$ we find that the deflection angle becomes
\begin{widetext}
\begin{eqnarray}\label{al3}
    \alpha&=& -3 M \sigma u_{n}^{\gamma -1}\int_0^1 \frac{ (x^2+x+1) (x^{\gamma }-1)}{ (1-x)^{3/2} (x+1)^{5/2}} \, dx+u_{n}^2 \Big[\Big(\frac{15 \pi }{4}-4\Big) M^2-\frac{3 \pi  Q^2}{4}\Big]\nonumber\\
    &&-M (u_{o}^2+u_{s}^2)(M+u_n^{-1})+4 M u_{n}+\Big[\frac{\sqrt{\pi }\, \Gamma \big(\frac{\gamma +1}{2}\big) u_{n}^{\gamma -2}}{ \,\Gamma \big(\frac{\gamma }{2}\big)}-\frac{1}{2} (u_{o}^{\gamma -1}+  u_{s}^{\gamma -1})u_n^{-1}\Big]\sigma  \,.
\end{eqnarray}
\end{widetext}

We see from~\eqref{al1}-\eqref{al3}, as was the case with~(\ref{s1}) corresponding to $\gamma =1$, that the corrections to the deflection angle for the Reissner-Nordstr\"om black hole (in the absence of quintessence) when the observer and the source are at large, but finite, distances ($r_o=1/u_o,r_s=1/u_s$) from the lens do not depend on the charge up to $u_o^2$ and $u_s^2$. All these corrections do not depend on $\sigma$ and are symmetric functions of ($u_o,u_s$), so they are easily recognized in Eqs.~\eqref{al1} to~\eqref{al3} and (\ref{s1}). Corrections due to quintessence are all functions of $\sigma$. Setting $\sigma =0$ in any one of the equations~\eqref{al1} to~\eqref{al3} and (\ref{s1}) yields the deflection angle for the Reissner-Nordstr\"om black hole.

All integrals over $x$ in Eqs.~\eqref{al1} to~\eqref{al3} do converge and could be given in closed forms, however, for some values of $\gamma$ only. For instance, for $\gamma =3/2$ the integral in~\eqref{al3} is given in terms of the complete elliptic integral $E(m)$ and the complete elliptic integral of the first kind $K(m)$
\begin{multline*}
\int_0^1 \frac{(x^2+x+1)(x^{3/2}-1)}{ (1-x)^{3/2} (x+1)^{5/2}} \, dx=\frac{2}{3}-\frac{7E(1/2)}{\sqrt{2}}\\+\frac{5\sqrt{2}K(1/2)}{3}.
\end{multline*}

How quintessence affects the deflection angle can be seen from the coefficient $C_{\sigma}$ of $\sigma$ in Eqs.~\eqref{al1}-\eqref{al3}, and~(\ref{s1}). For instance, in~\eqref{al3} we have
\begin{equation}\label{Cs}
C_{\sigma}\equiv \frac{\sqrt{\pi }\, \Gamma \big(\frac{\gamma +1}{2}\big) u_{n}^{\gamma -2}}{ \,\Gamma \big(\frac{\gamma }{2}\big)}-\frac{1}{2} (u_{o}^{\gamma -1}+  u_{s}^{\gamma -1})u_n^{-1}.
\end{equation}
For fixed ($u_0,\,u_s,\,u_n$) satisfying~\eqref{m6}, the coefficient $C_{\sigma}$ has a smooth variation for $1\leq\gamma <2$. This follows from the series expansions of $C_{\sigma}$ in the vicinity of $\gamma=2$ and $\gamma=1$, respectively:
\begin{align}
\label{al4}&C_{\sigma}=\frac{\pi}{2}-\frac{1}{2 u_n}(u_o+u_s)+\mathcal{O}(\gamma-2),\\
\label{al5}&C_{\sigma}=-\frac{1}{2 u_n}~\ln \Big(\frac{u_o u_s}{4 u_n^2}\Big)(\gamma -1)+\mathcal{O}(\gamma-1)^2.
\end{align}
By~\eqref{m6}, the second term in~\eqref{al4} is neglected with respect to the first term, so the coefficient $C_{\sigma}$ varies roughly between $0$~\eqref{al5} and some factor of $\pi$ for $1\leq\gamma <2$. Thus, for $\gamma$ larger than unity, the effect of quintessence almost drops and the values of the deflection angle are not very sensitive to variations in the values of $\gamma$.

\section{Conclusion}

The motion of photons around black holes is one of the most studied
problems in black hole physics. The behavior of light near black
holes is important to study the structure of spacetime near black
holes. In particular, if the light returns after circling around the
black hole to the observer, it cause a gravitational lens
phenomenon. Light passing by the black hole will be deflected by
angle which can be large or small depending on its distance from the
black hole.

In present paper, we have extended our previous work for the Kiselev
black hole by including the effects of the electric charge. This
extra parameter enriches the structure of spacetime with an
additional horizon. By solving the geodesic equations, we have
obtained the null geodesic structure for this black hole. Moreover
the lens equation provides the information about the bending angle.
For a general $w_q$, we managed to find an analytical expression for
the bending angle in the strong deflection limit considering
quintessence as a perturbation to the Reissner-Nordstr\"{o}m. Since
this geometry is non asymptotically flat, one needs to be very
careful to compute the bending angle since  the standard approach,
i.e. using the bending formula (See \cite{weinb}), cannot be applied
any more.

Instead of this approach, by using perturbation techniques and
series expansions (assuming some physical conditions on the
parameters), we directly integrate Eq. (\ref{e6a}) for all $w_q$ to
find the bending angle. The final expression of the bending angle in
the strong limit Eqs.~\eqref{al1} to~\eqref{al3} and (\ref{s1})
contain some corrections to the deflection angle obtained by a
Reissner-Nordstr\"{o}m black hole, which are proportional to the
normalization parameter $\sigma$, as well as corrections due to the
finiteness of the distances of the source and observer to the lens.

It is instructive to compare the results of deflection in the presence of quintessence with those in the presence of phantom fields. In Ref.~\cite{paths} light paths of normal and phantom Einstein-Maxwell-dilaton black holes have been investigated. It was emphasized that, in the presence of phantom fields, light rays are more deflected than in the normal case. Adopting the Bozza's formalism~\cite{boz}, the authors of Ref.~\cite{ding} have shown that the lensing properties of the
phantom field black hole are quite similar to that of the
electrically charged Reissner-Norstr\"{o}m black hole, i.e., the
deflection angle and angular separation increase with the phantom
constant. A similar approach was adopted in \cite{eiroa} to study
lensing by a regular phantom black hole. These authors have demonstrated
that the deflection angle does not depend on the phantom field
parameter in the weak field limit, whereas the strong deflection
limit coefficients are slightly different form that of Schwarzschild
black hole (see also \cite{ste}). In our case, $C_{\sigma}$~\eqref{Cs} is
positive for $1<\gamma<2$. This means that the deflection angle is a bit larger if quintessence is present.

As a future work, one can also study the lensing for other
interesting configurations such as ``Nariai BHs", ``ultra cold BHs"
and also for rotating black holes surrounded by quintessence matter.
This type of work might be important to study the highly redshifted
galaxies, quasars, supermassive black holes, exoplanets and dark
matter candidates, etc.

\begin{acknowledgments}
S.B. is supported by the Comisi{\'o}n Nacional de Investigaci{\'o}n
Cient{\'{\i}}fica y Tecnol{\'o}gica (Becas Chile Grant
No.~72150066). The authors would like  to thank Azka Younas for useful discussions and her initial works in this work.
\end{acknowledgments}

\end{document}